\documentclass{iaus}
\usepackage{graphicx}

\title[Natal Super Star Clusters] 
{Probing the Birth of Super Star Clusters: Implications for Massive 
Star Formation}

\author[K.E. Johnson]   
{Kelsey E. Johnson$^{1,2}$%
\affiliation{$^1$Department of Astronomy, University of 
Virginia,Charlottesville, VA, 22903 USA \break $^2$ Hubble Fellow
\break email: kej7a@virginia.edu\\[\affilskip]}}

\pubyear{2005}
\volume{227}  
\pagerange{0}
\date{?? and in revised form ??}
\jname{Proceedings Title IAU Symposium}
\editors{A.C. Editor, B.D. Editor \& C.E. Editor, eds.}
\begin{document}

\maketitle

\begin{abstract}

Super Star Clusters are one of the most extreme star forming
environments in the universe, and the most massive and dense of these
may be proto globular clusters.  Like individual massive stars, the
earliest stages of super star cluster evolution are deeply obscured,
and therefore our knowledge about their birth environments is
currently very incomplete.  However, the study of natal super star
clusters has become somewhat of a cottage industry in recent years,
and the sample of such objects has been growing rapidly with
high-quality long-wavelength data now available from a number of
observatories.  The natal super star clusters identified in
thermal-infrared and radio observations represent the youngest stage
of massive star cluster evolution yet observed.  Their properties
appear to be similar to those of ultracompact H{\sc ii} regions in the
Milky Way, but scaled up in total mass and luminosity.  I will overview
what we think we know about these objects based on existing
observations, discuss their relationship to ultracompact H{\sc ii} regions,
present new models of their spectral energy distributions based on 3-D
simulations, and outline some of the most significant gaps in our
current understanding.  \keywords{stars: formation, H{\sc ii} Regions,
extinction, galaxies: star clusters, galaxies: starburst}
\end{abstract}

\firstsection 

\section{Introduction}

We know that nearly all massive stars form in clusters (see, e.g. de
Wit, these proceedings).  Therefore, if we wish to understand massive
star formation in general, we must understand the clustered mode of
star formation.  Super star clusters (SSCs) are the most extreme type
of stellar cluster (and also one of the most extreme star forming
environments in the universe), with stellar densities exceeding $10^4$
stars pc$^{-3}$ in their cores.  As ``extreme'' objects, one might be
tempted to simply think of them as cosmic curiosities.  However, these
objects are not only fascinating because of their extreme nature, but
also because they hold clues to an important mode of star formation
during the time of galaxy assembly.  Specifically, most present-day
research strongly supports the idea that SSCs are the adolescent
precursors to the ancient globular clusters that are ubiquitous around
massive galaxies in the local universe today.  However, one should
also bear in mind that the infant mortality rate of SSCs in likely to
be quite high \cite[(possibly as high as 99\%, Fall \& Zhang
2001)]{fall01}, and only the most robust of these objects will have
survived to the ripe old age of present-day globular clusters.  The
low survival rate of SSCs suggests that they must have been formed
prodigiously in the early universe, making this mode of star formation
even more important for understanding the early evolution of of today's
massive galaxies.

SSCs are also important components of galaxies in the local universe.
These clusters can have an incredible impact on their surrounding
interstellar medium (ISM), and in some cases even the intergalactic
medium (IGM).  Each of these clusters can host up to thousands of
massive stars that have powerful stellar winds, and these stars will
all die in a violent way within a few million years of each other.
The collective effect of these massive stars (both in their lives and
deaths!) can cause tremendous outflows, ionize large volumes of the
ISM and IGM, and enrich vast amounts of interstellar material
\cite[(e.g. Johnson et al. 2000, Heckman 2001, Martin, Kobulnicky, \&
Heckman 2002)]{johnson00,heckman01,martin02}

Moreover, SSCs are also important tools for probing the universe.
They are the most luminous type of ``simple stellar population'' (the
constituent stars share the same age and metallicity), which makes
them very useful ``test particles'' for investigating local galaxy
properties.  In addition, SSCs will be among the brightest point
sources visible in galaxies out to large distances, which makes them
very attractive targets for observations that must worry about surface
brightness issues.

\section{The Foundations of Super Star Cluster Research}
\label{history}

Research on SSCs is rapidly approaching a crossroads, and it is
therefore worth briefly reviewing the tremendous advances that have
been made in this field over roughly the last decade.  Perhaps most
importantly, we have learned that globular cluster formation continues
today.  This discovery came to the forefront of cluster research with
the {\it Hubble Space Telescope}, and caused a major revision of the
prevailing theories of globular cluster formation and galaxy
evolution.  In particular, the previously accepted theories proposed
that globular clusters were among the first bound structures formed in
the universe, formed directly out of primordial material
\cite[(e.g. Peebles \& Dicke 1968)]{peebles68}.  This idea was
particularly compelling because of the similarity between the
predicted Jean's mass for primordial conditions, and the observed mass
of globular clusters.  The fact that globular clusters continue to
form at the present time provides us with a window into the earlier
universe when their formation was prevalent.

Not only did the {\it Hubble Space Telescope} spark the flurry of
research in this area, but it has also been the work-horse; the vast
majority of super star cluster studies have utilized {\it Hubble} in
some way.\footnote{See e.g. the proceedings for \underline{The
Formation and Evolution of Massive Young Star Clusters}, Lamers,
Smith, \& Nota eds, 2003} Because of {\it Hubble} there is a sample of
globular clusters spanning almost every galaxy type, with a range of
ages and metallicities.  Large statistical samples of SSCs have been
compiled, and their luminosity function (and presumably mass function)
is well constrained (a power-law of $-2$) and roughly universal.  The
effective radii of many SSCs have been directly measured (typically $\sim
5$~pc), and these radii in turn provide constraints on cluster dynamics
and evolution.  These are only a few of the many results.

The advances in this field due to {\it Hubble} have been truly
remarkable.  However, at the time of this writing, there are no
missions in the foreseeable future that will have capabilities
comparable to {\it Hubble} in the optical wavelength regime.  Given
this prognosis, it is clear that further progress in SSC research will
depend critically on utilizing other tools.  With new
observational capabilities becoming available in the infrared to radio
wavelength regimes, it is likely that the next decade of SSC research
will be dominated by studying their earliest stages of evolution,
which these long wavelengths are particularly well-suited to probing.  For
this reason, understanding the details of individual massive star
formation in the Milky Way has become all the more important for
extragalactic research.

\begin{figure}[!]
\centerline{
\resizebox{5.3in}{!}{
\includegraphics{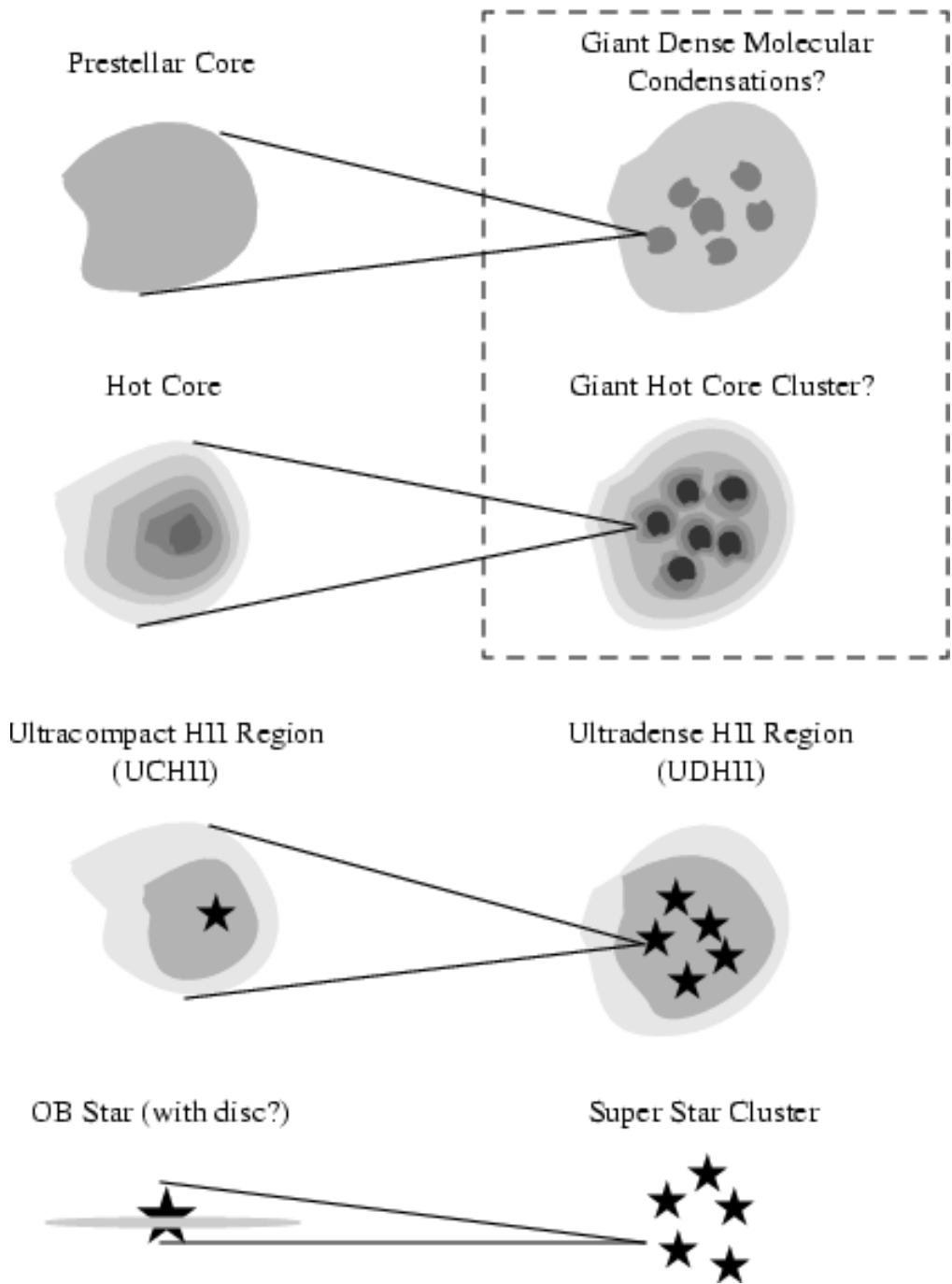}}}
\caption{A cartoon illustrating a parallel evolutionary sequence for 
individual massive stars and super star clusters.  The dashed box 
identifies the stages of super star cluster evolution that have not been 
observationally identified.}\label{cartoon}
\end{figure}

\section{Connections to Galactic Star Forming Regions}

There is, of course, a rich history of studying star formation in our
own Galaxy, as is highlighted in much of this volume.  While there are
clearly many important issues remaining in regard to massive star
formation, a great deal has also been learned.  Much of this knowledge
can be applied to extragalactic objects, which cannot be observed with as 
much detail and precision as their galactic counterparts.  

Observations of natal SSCs suggest that the early stages of massive
star cluster evolution may parallel those of individual massive stars
observed in the Milky Way.  In particular, natal SSCs appear to have
spectral energy distributions similar to ultra compact H{\sc ii}
regions; these infant clusters typically have been identified at radio
wavelengths as compact thermal (often optically thick) free-free
sources, and they have copious amounts of thermal infrared emission.
Like individual massive stars, SSCs also appear to spend $\sim
10-20$\% of the massive star lifetimes in a stage analogus to ultra
compact H{\sc ii} regions \cite[(Kobulnicky \& Johnson 1999, Vacca, Johnson, 
\& Conti 2002)]{kj99,vacca02}.  These similarities suggest that we can
apply many of the tools, techniques, and ideas regarding massive star
formation in our own galaxy to objects that are much more distant.
Figure~\ref{cartoon} illustrates an analogy between the evolution of
an individual massive star and a SSC.

While the analogy between individual massive stars and SSCs is a good
starting point, the extent to which this analogy holds true will
depend on a variety of factors.  For example, if star formation within
a SSC is not ``instantaneous'' to within $< 1$~Myr, then the
constituent stars in a cluster will not all be in the ultracompact
H{\sc ii} region stage simultaneously, which may result in different
global SEDs.  Extrapolating from what we know about individual massive
stars to SSCs could also get us into trouble because of the extreme
proto-stellar densities present in natal SSCs.  It seems likely that
we will gain the most physical insight not necessarily from the
extreme cases alone, but also from trying to understand how the
properties of star formation scale between individual massive star
formation and SSC formation.

\section{What We Think We Know About Natal Super Star Clusters 
\label{properties}}

To date, the attempts to constrain the physical properties of natal
clusters have been fairly crude, largely due to the limited
observations available.  As a result, we are forced by Occam's razor
to the most simple solution, one which requires the least free
parameters.  Our current concept is the following: When an SSC is
born, the massive stars ionize the surrounding natal material,
creating an extremely dense H{\sc ii} region, similar to a galactic
ultracompact H{\sc ii} region, but scaled up in mass and luminosity.
This dense H{\sc ii} region is in turn surrounded by a dust cocoon
that is presumably quite warm on the inner boundary.  The constituent
ionizing stars may be surrounded by individual H{\sc ii} regions and
dust cocoons in the earliest stages, but this is currently unknown.
Clusters that are more sparse are likely to resemble ultracompact
H{\sc ii} region complexes in the Milky Way (such as W49A), and
contain spatially discrete H{\sc ii} regions and cocoons, but the
existence of individual cocoons becomes less tenable in the extreme
stellar densities of the cores of massive SSCs.

These components of a natal cluster have separate observational
signatures. The dense H{\sc ii} regions are characterized by an
``inverted'' radio spectral energy distributions due to optically
thick free-free emission ($\alpha >0$, where $S_\nu \propto
\nu^\alpha$).  The frequency at which the radio emission becomes
optically thick is largely dependent on the density of the H{\sc ii}
region.  In the earliest stages of cluster evolution, most or all of
the stellar luminosity is reprocessed by the surrounding dust cocoon,
and this cocoon is observable at thermal infrared to sub-millimeter
wavelengths.  The nature of the resulting spectral energy distribution
depends on whether the dust in the cocoon is smooth or clumpy (as will
be discussed in \S~\ref{models}, but this is not currently constrained
by observations.

Existing observations have provided {\it estimates} for some of the
physical properties of natal clusters, including size, electron
density, pressure, stellar mass, H{\sc ii} mass, dust mass, and age.
For the most nearby clusters, their size can be directly measured
using high spatial resolution observations.  For more distant
clusters, model H{\sc ii} regions can be used to fit the radio
spectral energy distribution and infer both the radius and electron
density.  The embedded SSCs that have been analyzed in this way
\cite[(e.g. those in He~2-10, Johnson \& Kobulnicky 2003)]{jk03} have
H{\sc ii} region radii of only a few parsecs, and the electron
densities have global values of $\sim 10^3$~cm$^{-3}$, with peak
values in excess of $\sim 10^6$~cm$^{-3}$.  Assuming temperatures of
$\sim 10^4$~K, these densities imply global pressures of $P/k_B >
10^7$~K~cm$^{-3}$ and peak pressures reaching values of $P/k_B \sim
10^{10}$~K~cm$^{-3}$.

The masses of the natal clusters, H{\sc ii} regions, and dust cocoons
can be inferred from radio and infrared observations.  The
optically-thin thermal flux (typically measured with the highest
available radio frequency) can be used to infer a total ionizing
luminosity, and the ionizing luminosity can be translated into an
embedded stellar mass by assuming a stellar initial mass function.
Using this technique, clusters with a range of masses have been
detected (spanning modest OB-associations to massive SSCs).  The H{\sc
ii} mass can be estimated given the radii and densities for the
ionized regions.  In at least some cases, the H{\sc ii} mass has
anomalously low values \cite[($< 5$\% of the stellar mass, Johnson \&
Kobulnicky 2003)]{jk03}, which we tentatively interpret as a sign of
the youth of the H{\sc ii} regions.  The mass of the dust cocoons
around embedded SSCs has also been estimated in a few cases using
infrared observations; these estimates have suggested dust masses of
$\sim 10^5 - 10^6 M_\odot$ for massive SSCs in SBS0335-052 and He~2-10
\cite[(Plante \& Sauvage 2002, Vacca, Johnson, \& Conti
2002)]{plante02, vacca02}.

\section{An Important Case Study: Henize 2-10}

A critical and unresolved issue in the study of massive star clusters
is whether SSCs are just the statistical tail of a continuous mass
distribution that extends all the way down to individual massive
stars, or whether they can only be formed in special environments.
While there is some evidence that supports each side of this debate,
the answer has been clouded by observational and evolutionary issues.
In typical optical studies, a cluster's age, extinction, and mass are
all contributing factors to its observed luminosity and inferred mass.
Moreover, as clusters evolve, the lower mass clusters will tend to be
disrupted on time-scales as short as a few million years
\cite[(Portegies Zwart et al. 2002)]{pz02}, causing further confusion
when interpreting cluster luminosity and mass functions.  Radio
observations of natal clusters are immune to reddening and sample only
the youngest clusters before they have had time to evolve.  As a
result, only radio observations are capable of directly sampling the
cluster mass function at the time of their birth.

The starburst galaxy Henize 2-10 is one of the nearest galaxies
(9~Mpc) known to host multiple natal SSCs detected in radio and
infrared observations, as well as numerous SSCs that have already
emerged to be visible in optical light \cite[(Kobulnicky \& Johnson
1999, Johnson et al. 2000, Vacca, Johnson, \& Conti 2002, Johnson \&
Kobulnicky 2003, Cabanac, Vanzi, \& Sauvage
2005)]{kj99,johnson00,vacca02,jk03,cabanac05}.  These properties make
Henize~2-10 a very important case study for examining the properties
of natal SSCs.  In our previous radio observations, four natal SSCs
were detected, all with inferred masses in the range of $\sim 1-5
\times 10^5 M_\odot$.  Although the observations were sensitive enough
to detect SSCs with somewhat lower masses, less massive natal clusters
masses were not apparent. Given the linear resolution of the
observations used for these previous radio studies of $\sim 20$~pc, it
seemed to be likely that some lower mass clusters were missing due to
crowding and confusion.

\begin{figure}[!]
\centerline{
\resizebox{5.5in}{!}{
\includegraphics{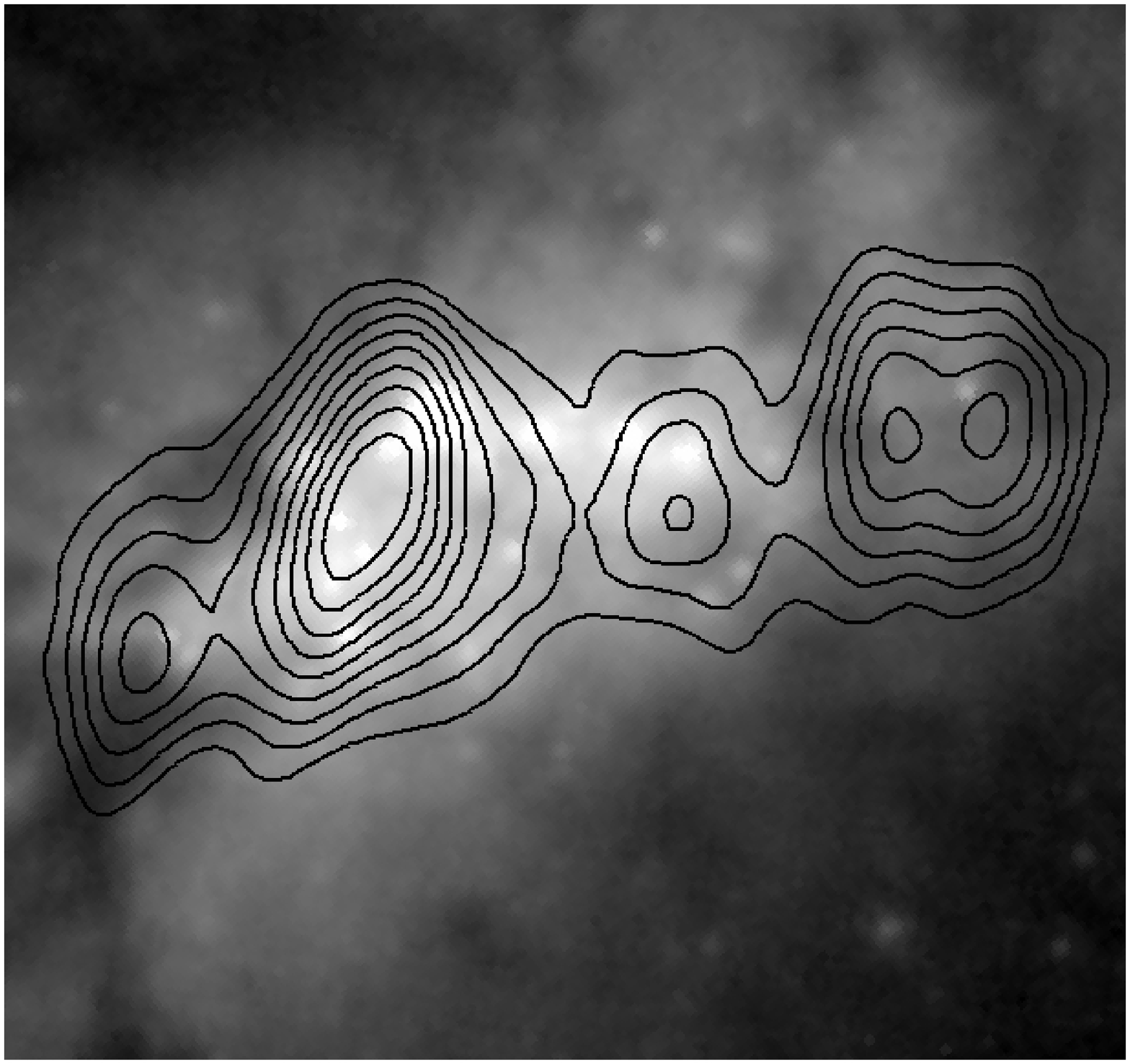} \includegraphics{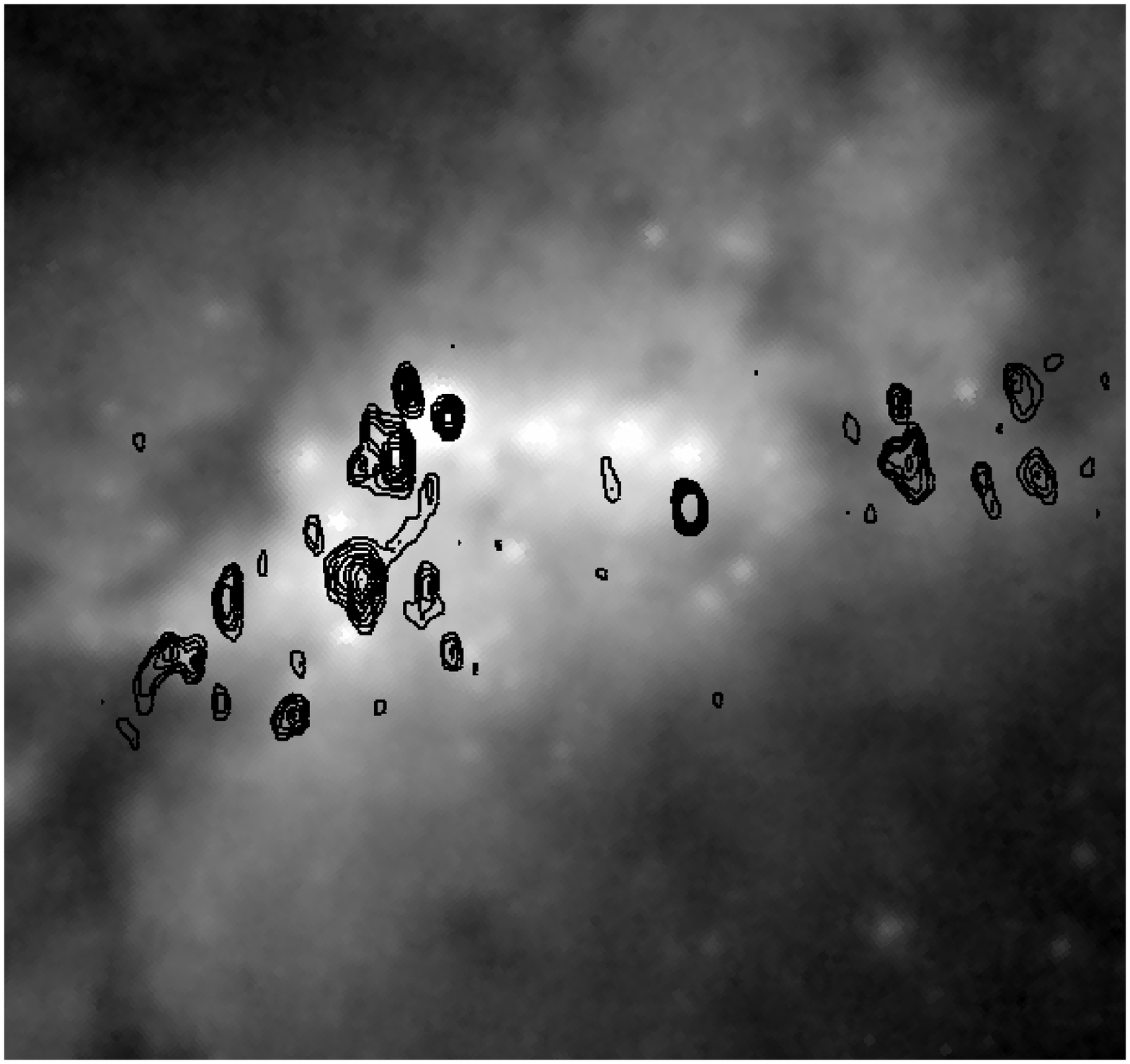}}}
\caption{A comparison between the old (left) and new (right) radio
observations from the Very Large Array (shown in contour).  The
gray-scale image in each panel is an I-band image from {\it HST}.  The
previous radio observations had a linear resolution of $\sim 20$~pc
\cite[(Kobulnicky \& Johnson 1999, Johnson \& Kobulnicky 2003)]{kj99,
jk03}, while the new observations that include the Pie Town antenna have
a linear resolution of $\sim 4$~pc \cite[(Biswas \& Johnson in
prep)]{bj03}.  It is clear that the new radio observations will shed
light on many of the issues related to SSC formation. }\label{HE210}
\end{figure}

In order to address the question of whether Henize~2-10 is only
forming SSCs with masses of $\sim 10^5 M_\odot$, or whether lower mass
clusters are present, but had not yet been detected or disentangled
from the more luminous sources, we obtained new high-resolution
observations with the Very Large Array using the additional Pie
Town antenna.  A comparison between the old and new radio observations
is shown in Figure~\ref{HE210} \cite[(Biswas \& Johnson in
prep)]{bj05}. In the new observations, it is clear that many of the
previously known radio sources are composed of multiple components.
Although these observations are also roughly an order of magnitude
more sensitive than the previous observations, it is not clear that
any additional natal SSCs exist outside of the main star forming
region apparent in the previous radio observations.  The analysis
is still underway at the time of this writing, but it is clear that
these new high-resolution and extremely sensitive radio observations
will shed light on many issues related to SSC formation.

\section{The Infrared Properties of Natal SSCs: Results from 3-D 
Monte Carlo Radiation Transfer Models \label{models}}

With the {\it Spitzer Space Telescope} providing a wealth of new
observations, and {\it Herschel}, {\it SOFIA}, and ALMA on the
horizon, appropriate physical models of natal SSCs are required in
order to analyze data from these facilities.  While the simple
conceptual picture discussed in \S~\ref{properties} is a good place to
start, we know that the universe is complicated, and sometimes nice
well-behaved uniform spheres just will not fit the data.  One of the
main difficulties that must be overcome in order to properly model
star forming environments is that the ISM is not smooth, and therefore
1-D models cannot reproduce physical structures that are consistent
with the observed ISM.  We have produced a suite of models of natal
SSCs using the 3-D Monte Carlo radiation transfer code described by
\cite[Whitney et al. (2003)]{whitney03}.  These models allow for
arbitrary density distributions, and we have invoked a clumpy fractal
structure, consistent with observations of the ISM.

\begin{figure}[!]
\centerline{
\resizebox{4.5in}{!}{
\includegraphics{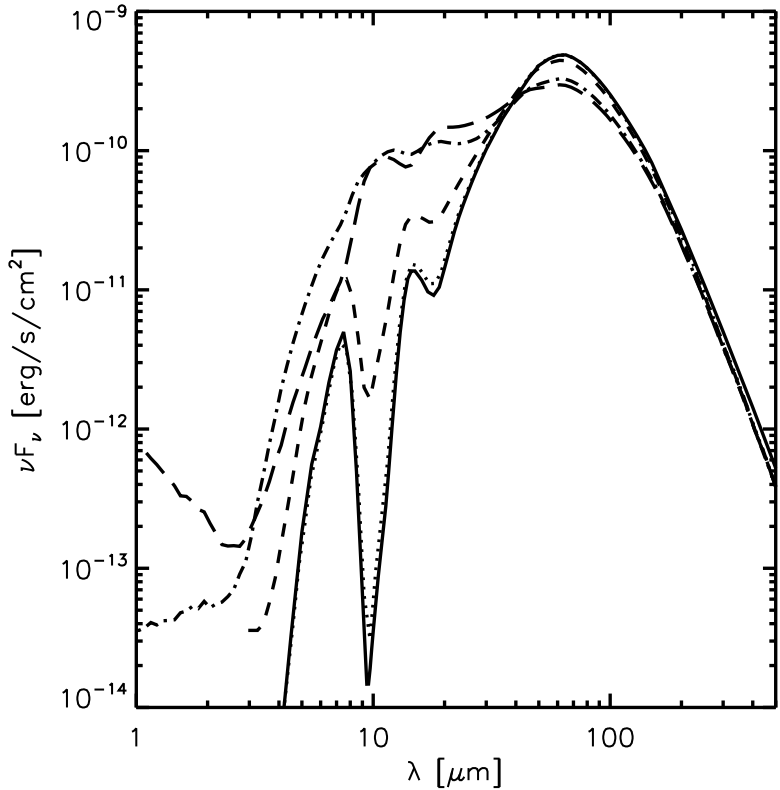}}}
\caption{A comparison between the modeled spectral energy distribution for 
a natal SSC using different fractions of clumpy dust.  The models shown 
are for a $10^6 M_\odot$ stellar cluster formed with 10\% star formation 
efficiency, an outer dust cocoon radius of 25~pc, and an inner cocoon 
radius of 1~pc.  The solid line is 100\% smooth dust, dotted line is 90\% 
smooth, short-dashed is 50\% smooth, long-dashed is 10\% smooth, and 
dot-dashed is 1\% smooth. Note that this figure is shown on a log-log scale, 
and the differences between the models can be quite significant. }\label{SED}
\end{figure}

The differences in the spectral energy distribution (SED) between
smooth and clumpy models can be very significant, as shown in
Figure~\ref{SED}.  One of the main differences is that clumpy dust
allows more near-infrared light (both scattered and direct from the
central source) to escape.  More reprocessing of the photons also
occurs in smooth dust cocoons, which changes the shape of the
resulting SED.  Spectral features, such as the silicate line at $\sim
10\mu$m can also vary dramatically depending on the clumpy fraction
for model with otherwise identical parameters (the silicate feature is
much more strongly absorbed in models with a larger smooth fraction).

One important result is that the mid-infrared SED of the same object
can vary significantly depending on the viewing angle if the dust
cocoon is moderately clumpy.  This is very bad news from an
observational perspective; this degeneracy makes it impossible,
regardless of the number of data points that sample a given SED, to
determine a unique set of physical parameters that will fit the
observations.  The good news is that the degeneracy introduced by
clumping is less significant at far-infrared wavelengths, although
these wavelengths also suffer from poor spatial resolution.

These results should serve as a strong warning that cannot be
overstated to investigators with thermal-infrared observations: {\it
Using 1-D models could lead to a significant misinterpretation of the
data.}

There is a vast amount of parameter space to explore, even without
allowing for 3-D structure.  The current study is limited to
investigating the effects of a varying the inner and outer radius of
the dust cocoon, total initial dust mass of the cocoon (which directly
reflects the star formation efficiency), and the fraction of the dust
that is clumpy.  Within this limited parameter space, we have
simulated an evolutionary sequence for SSCs as they go from deeply
embedded to only having a thin dust shell.  The results from this
evolutionary sequence can be directly compared to observed colors of
ultra compact H{\sc ii} regions.  

\begin{figure}[!]
\centerline{
\resizebox{4.5in}{!}{
\includegraphics{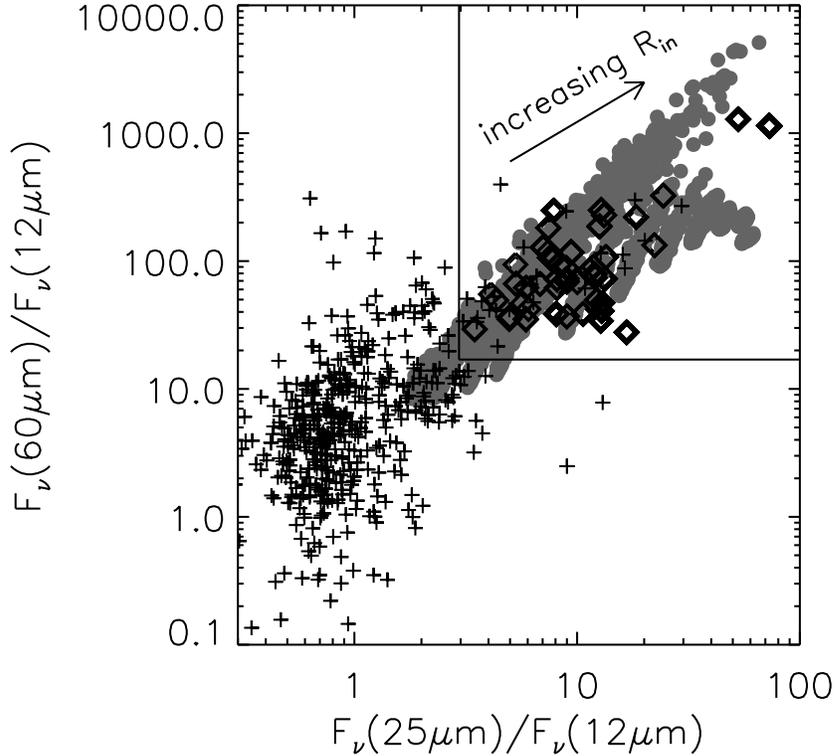}}}
\caption{The {\it IRAS} colors from a subset of our models (gray
circles) are compared to field objects (plus signs) and ultra compact
H{\sc ii} regions (diamonds) from the Kurtz, Churchwell, \& Wood
(1994) survey.  The models shown are for a $10^6 M_\odot$ stellar
cluster formed with 10\% star formation efficiency, and an outer dust
cocoon radius of 25~pc.  The inner radius evolves through our
pseudo-evolutionary sequence and moves outward from 0.1~pc to 24~pc.  All
of the sight lines are shown, resulting in the scatter.  The box shown
in the upper right is the color criteria defined by Wood \& Churchwell
(1989) to select ultra compact H{\sc ii} regions.  }\label{WC}
\end{figure}

Figure~\ref{WC} shows a comparison between the colors of natal SSCs
from our pseudo-evolutionary sequence produced by our 3-D models
and the observed {\it IRAS} colors of ultra compact H{\sc ii} regions
as well as ``field'' objects \cite[(Johnson et al. in prep, Kurtz,
Churchwell, \& Wood 1994)]{johnson05,kurtz94}.  Overall, there is
excellent agreement between our model colors and the observed colors
of ultra compact H{\sc ii} regions.  In general, as a natal SSC moves
through our pseudo-evolutionary sequence (its inner radius becomes
larger and therefore the dust becomes cooler), its {\it IRAS} colors
become more red.  The most deeply embedded objects lie in the lower
left region of the spread of model points, which reflects the
extremely hot dust in the inner portions of the dust cocoons.  If
these objects with such hot dust exist, the lifetime of this state is
likely to be extremely short.  Therefore, embedded star forming
regions with such hot dust should be quite rare, possibly accounting
for the lack of ultra compact H{\sc ii} regions detected with colors
as blue as the models can produce.  The model colors also extend
further red-ward than observed for ultra compact H{\sc ii} regions, but
this is not surprising; as an embedded star forming region evolves and
begins to emerge from its cocoon, its thermal-infrared luminosity will
fall precipitously.  As a result, we expect that the existing samples
of ultra compact H{\sc ii} regions become incomplete at these later
stages (or perhaps objects that are this evolved are no longer
considered to be ultra compact H{\sc ii} regions!).

\section{Some Major Unresolved Questions}
There are no lack of questions regarding the birth of SSCs, and given
the new observatories that are becoming available, it is an
exceptionally good time to be asking these questions.  A few major
questions keep recurring that I want to highlight here:

(1) How does star formation vary between individual massive stars 
and SSCs (and is this reflected in the stellar initial mass
function)?

(2) What are the environmental requirements necessary to form a SSC, and 
how do they differ from individual stars?

There are also a variety of properties that it would be nice to know
more about.  For example, it would be instructive to {\it directly}
measure densities, pressures, and temperatures using various line
diagnostics at IR to radio wavelengths.  Another quantity that might
be particularly useful for theorists (that is almost completely
unconstrained) is the star formation efficiency in these objects.  We
also have no idea how much ionizing radiation could be escaping from
the natal cocoons (and it could be quite a lot if the dust is clumpy).
In fact, we know almost nothing about the natal dust cocoons
(including the temperatures, grain compositions, and geometry).  This
list is certainly not exhaustive.  An understanding of these
quantities, and how they relate to the properties of the resulting
cluster, will eventually provide a great deal of insight into the
requirements for massive star cluster formation.

\section{Looking Forward to the Future}
\begin{figure}[!]
\centerline{
\resizebox{4.5in}{!}{
\includegraphics{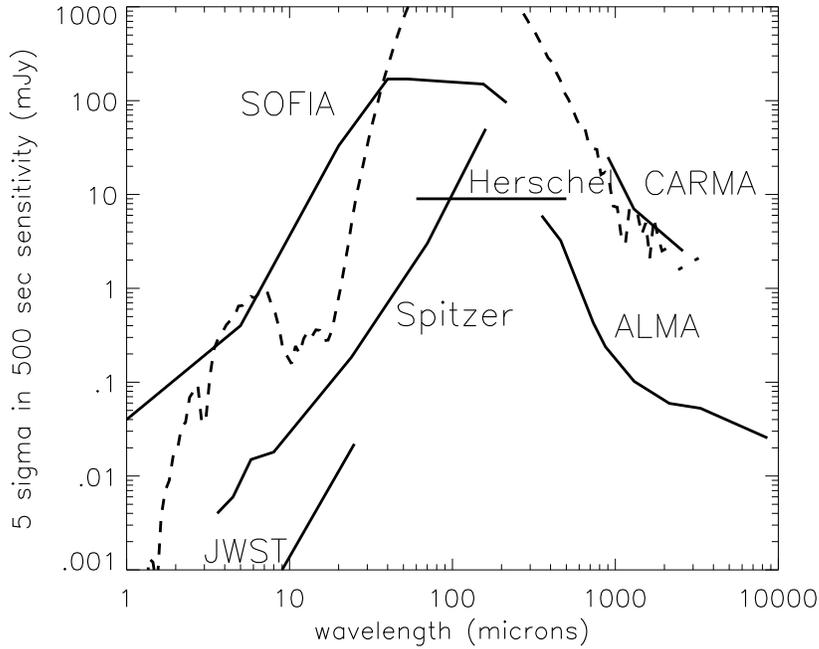}}}
\caption{The anticipated sensitivities for upcoming infrared and
millimeter observatories normalized to 500~sec of integration time.  A
sample model spectral energy distribution for a $10^6 M_\odot$ natal
SSC at a distance of 10~Mpc is over-plotted (dashed line) for
comparison (Johnson et al. in prep).  These new observing facilities
are extremely well-suited to probing the birth of SSCs. }\label{obs}
\end{figure}

Over the next decade (or so), a number of powerful new facilities are
going to become available at infrared to radio wavelengths.  These
observational capabilities are likely to bring about a renaissance in
the study of the formation of massive clusters.  These facilities will
be incredibly well-suited to observing natal SSCs (Figure~\ref{obs}),
and will allow us to study natal SSCs in amazing detail.  Surely the
new observations will bring about many new discoveries as well as many
more new questions.

\begin{acknowledgments}
The work has benefited from many fruitful discussions with my
collaborators, including I. Biswas, P. Conti, M. Goss, L. Hunt,
R. Indebetouw, H. Kobulnicky, W. Vacca, B. Whitney, and K. Wood.  I
also gratefully acknowledge support for this work provided by the NSF
through and Astronomy and Astrophysics Postdoctoral Fellowship, and by
NASA through the Hubble Fellowship Grant \#01173.02-A through the
Space Telescope Science Institute, which is operated by the Associated
Universities for Research in Astronomy, Inc., under NASA contract
NAS5-26555.

\end{acknowledgments}

\end{document}